\documentstyle[tighten,twocolumn,aps,psfig]{revtex}
\draft
\begin{document}
\title{ Chiral and 
$U(1)_A$ restorations high in the hadron spectrum
and  the semiclassical approximation.}
\author{ L. Ya. Glozman}
\address{  Institute for Theoretical
Physics, University of Graz, Universit\"atsplatz 5, A-8010
Graz, Austria\footnote{e-mail: leonid.glozman@uni-graz.at}}
%\maketitle

\twocolumn[
    \begin{@twocolumnfalse}
      \maketitle
      \widetext
\begin{abstract} 
In quantum systems with large $n$ (radial quantum number) 
or large angular momentum the semiclassical (WKB)
approximation is valid. A physical content of the
semiclassical approximation is that the quantum
fluctuations effects are suppressed and vanish asymptotically.
The chiral as well as $U(1)_A$ breakings in QCD is a result
of quantum fluctuations. Hence these breakings must be
absent (suppressed) high in the spectrum and the spectrum of high-lying
hadrons must exhibit  symmetries of the classical QCD
Lagrangian.

\end{abstract}

  \end{@twocolumnfalse}
  ]
{
    \renewcommand{\thefootnote}%
      {\fnsymbol{footnote}}
    \footnotetext[1]{e-mail address: leonid.glozman@uni-graz.at}
    }
\narrowtext

\bigskip
\bigskip

If one neglects tiny masses of $u$ and $d$ quarks, which
are much smaller than $\Lambda_{QCD}$ or the typical
hadronic scale of 1 GeV, then the QCD Lagrangian
exhibits the

\begin{equation}
U(2)_L\times U(2)_R = SU(2)_L\times SU(2)_R\times U(1)_V\times U(1)_A
\label{sym}
\end{equation}

\noindent
symmetry. This is because the quark-gluon interaction
Lagrangian in the chiral limit does not mix the left- and
right-handed components of quarks and hence the total
QCD Lagrangian for the two-flavor QCD can be split into
the left-handed and right-handed parts which do not communicate
to each other. We know that the $U(1)_A$ symmetry of the
classical QCD Lagrangian is absent at the quantum level
because of the $U(1)_A$ anomaly \cite{ANOMALY}. We also know
that the chiral $SU(2)_L\times SU(2)_R$ symmetry is spontaneously
broken in the QCD vacuum. That this is so is directly evidenced
by the nonzero value of the quark condensate, 
$\langle \bar q q \rangle = \langle \bar q_L q_R + \bar q_R q_L \rangle
\simeq -(240 \pm 10 MeV )^3$, which represents an order
parameter for spontaneous chiral symmetry breaking. This
quark condensate directly shows that
in the QCD vacuum the left-handed quarks are correlated
with the right-handed antiquarks (and vice verca) and
hence the QCD vacuum breaks the chiral symmetry.\\

That the chiral symmetry is spontaneously broken is also
directly seen from the low-lying hadron spectrum. If the
chiral symmetry were intact in the vacuum, i.e. it were
realized in the Wigner-Weyl mode, then all hadrons would
fall into parity-chiral multiplets \cite{CG1}, i.e.
multiplets of the $SU(2)_L\times SU(2)_R \times C_i$ group,
where $C_i$ consists of identity and space inversion. In the
baryon spectrum these multiplets are either parity doublets
in $N$ and $\Delta$ spectrum that are not related to each other,
or quartets that contain degenerate parity doublets in the nucleon
and delta spectra with the same spin. From the low-lying nucleon
and delta spectra we definitely conclude 
that there are no degeneracies of states
of the same spin but opposite parity. This tells that chiral symmetry
must be broken in the QCD vacuum, i.e. it is realized in the Nambu-Goldstone
mode. Even more, there is no one-to-one mapping of the states
with the same spin and opposite parity low in the spectrum.
This suggests that low in the spectrum the chiral symmetry is not
only strongly broken, but in addition realized nonlinearly \cite{W}.
Similar, in the meson spectrum the unbroken chiral symmetry
would imply that e.g. $\pi$-mesons $(I,J^P=1,0^-)$ and pure
$n\bar n = \frac{u\bar u + d\bar d}{\sqrt 2}$ $f_0$ states
$(I,J^P=0,0^+)$ would be systematically degenerate, level by
level. Clearly it is not the case for the low-lying states (see Fig. 2).\\

The high-lying hadrons, however, show obvious signs of
parity doubling. For example, all the excited nucleons around
1.7 GeV are well established states and we see here three
approximate parity doublets with spins 1/2, 3/2 and 5/2. Similar,
the lowest excitations with $J=9/2$ are also well established
states and they represent another good example of parity
doubling. Not established states (i.e. candidates, that are
marked as "**" and "*" states according to PDG classification
\cite{PDG}) also support parity doubling, though the uncertainties
are very high (one should not take too seriously "*" states, of course).
There is a well established state with $J=11/2$, where so far
no parity partner has been seen. Similar situation occurs 
in the delta spectrum.
It has been suggested recently \cite{G1,CG1} that
this parity doubling reflects effective chiral symmetry
restoration in high-lying hadrons.\footnote{This phenomenon
has been refered to \cite{ERICE} as chiral symmetry restoration
of the second kind in order to distinguish it from the
chiral symmetry restoration in the QCD vacuum at high temperature
or density.} Unfortunately from
the baryon spectrum alone we cannot distinguish whether the parity
doublets evidence the restoration of chiral or $U(1)_A$ 
symmetries\footnote{The restoration of the $U(1)_A$ symmetry,
but not of the chiral symmetry, 
has been suggested by Jaffe as explanation of parity doublets
in the baryon spectrum, as cited in ref. \cite{CG1}. It has been 
shown in  \cite{CG1}, however,
that $U(1)_A$ cannot be restored without restoration of 
$SU(2)_L \times SU(2)_R$. This is because even if the effects of
the axial anomaly vanish, the $U(1)_A$ would be still 
broken if the $SU(2)_L \times SU(2)_R$ is broken.
This is because the same quark condensates in the QCD vacuum break
both types of symmetries.}.\\

The systematic data on high-lying mesons are still absent
in the PDG tables. The results of the ongoing partial wave 
analysis  \cite{BUGG1,BUGG2,BUGG0} of high-lying
mesons obtained in $p\bar p$ annihilation at LEAR
suggest a clear evidence \cite{G2,G3} for the chiral symmetry 
restoration. This is well seen from the Fig.2,
where the high-lying $\pi$ mesons and $n\bar n$ $f_0$ mesons
are shown. There are indications of simultaneous chiral and $U(1)_A$
restorations in highly excited mesons \cite{G2}, since the
highly excited $\pi$, $a_0$, and $n\bar n$
$f_0$ as well as $n\bar n$ $\eta$ mesons form approximately degenerate
multiplets of the $U(2)_L \times U(2)_R$ group.\\

\begin{figure}
\hspace*{-0.5cm}
\centerline{
\psfig{file=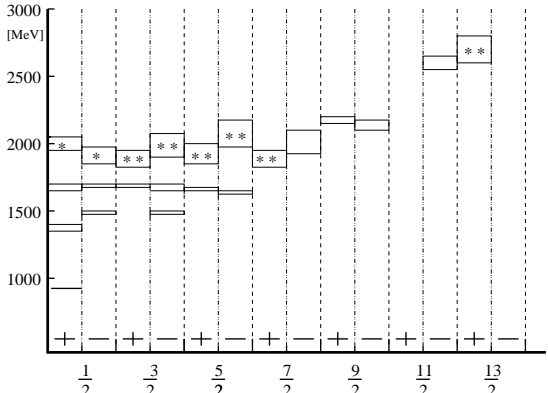,angle=-90,width=0.4\textwidth}
}
\caption{Excitation spectrum of the nucleon. The real part
of the pole position is shown. Boxes represent experimental
uncertainties. Those resonances which are not yet established
are marked by two or one stars according to the PDG classification.
The one-star resonances with $J=1/2$ around 2 GeV are given
according to the recent Bonn (SAPHIR) results. }
\end{figure}

By definition an
 effective symmetry restoration means the following. In QCD the
hadrons with the quantum numbers $\alpha$ are created
when one applies the local interpolating field (current) $J_\alpha(x)$
with such quantum numbers  on the vacuum 
$|0\rangle$. Then all the
hadrons that are created by the given interpolator
appear as intermediate states in the two-point correlator

\begin{equation}
\Pi_{J_\alpha}(q) = i \int d^4x~ e^{-iqx}
\langle 0 | T \left \{ J_\alpha(x) J_\alpha(0)\right \} | 0 \rangle,
\label{corr}
\end{equation}

\noindent
where all possible Lorentz and Dirac indices (which are specific for
a given interpolating field) have been omitted, for simplicity.
Consider two local interpolating fields  $J_1(x)$ and 
$J_2(x)$ which are connected by chiral transformation,
$J_1(x) = U J_2(x) U^\dagger$, where $U \in SU(2)_L \times SU(2)_R$
(or by $U(1)_A$ transformation). Then if the vacuum is
invariant under the given symmetry group, $U|0\rangle = |0\rangle$,
it follows from (\ref{corr}) that the spectra created by the
operators  $J_1(x)$ and  $J_2(x)$ must be identical.

\begin{figure}
\hspace*{-0.5cm}
\centerline{
\psfig{file=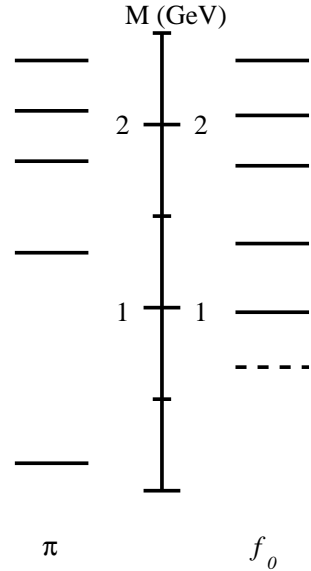,angle=-90,width=0.6\textwidth}
}
\caption{Pion and $n \bar n$ $f_0$ spectra. The three highest
states in both pion and $f_0$ spectra are taken from \protect\cite{BUGG1,BUGG2,BUGG0}.
Since  these $f_0$ states are obtained in $p \bar p$ and   they
decay predominantly  into $\pi \pi$ channel, they are considered in 
\protect\cite{BUGG1,BUGG2,BUGG0} as  $n \bar n$ states.}
\end{figure}

 We know that in QCD  
$U|0\rangle \neq |0\rangle$.
 As a consequence the corresponding spectral densities 
$\rho_1(s) \neq \rho_2(s)$. However,
it may happen that the noninvariance of the vacuum becomes
unimportant (irrelevant) high in the spectrum. Then the spectral
functions  $\rho_1(s)$ and $\rho_2(s)$ become close  at large $s$ 
(identical asymptotically high). This can be refered to as effective
chiral symmetry restoration from the low-lying spectrum, where
both $\rho_1(s)$ and $\rho_2(s)$ are very different because
of the symmetry breaking in the vacuum, to the high-lying spectrum,
where the asymmetry of the vacuum becomes unimportant and
$\rho_1(s)\approx \rho_2(s)$ (chiral symmetry restoration
of the second kind). We stress that this effective chiral symmetry
restoration does not mean that chiral symmetry breaking in
the vacuum disappears, but only that the role of the quark
condensates that break chiral symmetry in the vacuum becomes progressively
less important high in the spectrum \cite{CG1}. The valence
quarks in high-lying hadrons {\it decouple} from the QCD vacuum.\\

In ref. \cite{CG1} a justification for effective chiral symmetry
restoration  has been suggested. Namely,
at large space-like momenta
$Q^2 = -q^2 > 0$ the correlator
can be adequately represented by the operator product
expansion, where all nonperturbative effects reside in
different condensates \cite{SVZ}. The only effect that
spontaneous breaking of chiral symmetry can have on the
correlator is via the quark condensate of the vacuum,
$\langle \bar q  q \rangle$, and higher dimensional
condensates that are not invariant under chiral transformation $U$.
However, the contributions of all these condensates are suppressed
by inverse powers of momenta $Q^2$.  This shows that
at large space-like momenta the correlation function
becomes chirally symmetric. In other words

\begin{equation}
\Pi_{J_1}(Q) \rightarrow  \Pi_{J_2}(Q) ~~~ at ~~~Q^2 \rightarrow \infty.
\label{c0}
\end{equation}

\noindent
The dispersion relation provides a connection between the
space-like and time-like domains for the Lorentz scalar
(or pseudoscalar) parts of the correlator. In particular,
the large $Q^2$ correlator is completely dominated by the
large $s$ spectral density $\rho(s)$, which is an observable.
Hence the large $s$ spectral density should be insensitive
to the chiral symmetry breaking in the vacuum and must
satisfy 

\begin{equation}
\rho_1(s) \rightarrow  \rho_2(s) ~~~ at ~~~s \rightarrow \infty.
\label{c1}
\end{equation}

\noindent
This is in contrast to the low $s$ spectral densities 
$\rho_1(s)$ and $\rho_2(s)$, which are very different
because of the chiral symmetry breaking in the vacuum.\\

While the argument above on the asymptotic symmetry
properties of spectral functions is rather robust 
(it is based actually only on
the asymptotic freedom of QCD at large space-like momenta
and on the analyticity of the two-point correlator), it
is not clear whether it can be applied to the bound state 
systems, which the hadrons are. Indeed, it can happen that
the asymptotic symmetry restoration applies only to that
part of the spectrum, which is above the resonance region
 (i.e. where the current creates jets but not isolated hadrons). So
the question arises whether it is possible to prove (or at
least justify) the symmetry restoration in highly excited
{\it isolated} hadrons. We show below that both chiral
and $U(1)_A$ restorations in highly excited isolated hadrons {\it must}
be anticipated as a direct consequence of the semiclassical
approximation.\\

At large $n$ (radial quantum number) or at large angular
momentum $L$ we know that in quantum systems the {\it semiclassical}
approximation (WKB) {\it must} work. Physically this approximation
applies in these cases because the de Broglie wavelength of
particles in the system is small in comparison with the
scale that characterizes the given problem. In such a system
as a hadron the scale is given by the hadron size while the
wavelength of valence quarks is given by their momenta. Once
we go high in the spectrum the size of hadrons increases as well as
 the typical momentum of valence quarks.
This is why a highly excited hadron  can be described semiclassically
in terms of the underlying quark and gluon degrees of freedom.\\

A physical content of the semiclassical approximation is
most transparently given by the path integral. The contribution
of the given path to the path integral is regulated by the
action $S(q)$ along the  path $q(x,t)$

\begin{equation}
\sim e^{iS(q)/\hbar}.
\label{path}
\end{equation}

\noindent
The semiclassical approximation  applies when $S(q) \gg \hbar$.
In this case the whole amplitude (path integral) is dominated by
the classical path $q_{cl}$ (stationary point) and those paths that are infinitesimally
close to the classical path. All other paths that differ from
the classical one by an appreciable amount  do
not contribute. These latter paths would represent the quantum fluctuation
effects. In other words, in the semiclassical case the quantum
fluctuations effects are strongly suppressed and vanish asymptotically.\\

The $U(1)_A$ symmetry of the QCD Lagrangian is broken only due
to the quantum fluctuations effects. The $SU(2)_R \times SU(2)_L$
spontaneous (dynamical) breaking is also pure quantum effect
and is based upon quantum fluctuations. To see the latter we
remind the reader that most generally the chiral symmetry breaking
(i.e.the dynamical quark mass generation) is formulated via the
Schwinger-Dyson equation. It is not yet clear at all
which specific gluonic interactions are the most important ones as a kernel
of the Schwinger-Dyson equation (e.g. instantons \cite{SS}, or 
gluonic exchanges \cite{RWA}, or perhaps  other gluonic
interactions, or a combination of different interactions).
But in any case the quantum fluctuations effects of the quark
and gluon fields are
very strong in the low-lying hadrons and induce both 
chiral and $U(1)_A$ breakings. As a consequence we do not
observe any chiral or $U(1)_A$ multiplets low in the spectrum.
However,  if the quantum fluctuations effects
are absent or suppressed due to some reasons, then the 
dynamical mass of quarks must vanish as well as effects of the $U(1)_A$
anomaly.\\

We have just mentioned that in a quantum system with
large enough $n$ or  $L$ the quantum fluctuations must be
suppressed and vanish asymptotically. Then it follows 
that in such systems both
the chiral and $U(1)_A$ symmetries must be restored.
Hence at large hadron masses (i.e. with either large $n$ or
large $L$) we must observe symmetries of the classical
QCD Lagrangian. This is precisely what we see
phenomenologically. In the nucleon spectrum the doubling
appears either at large $n$ excitations of baryons with
the given small spin or in resonances of large spin. Similar
features persist in the delta spectrum. In the meson spectrum
the doubling is obvious for large $n$ excitations of small
spin mesons (see Fig. 2) and there are signs of doubling of large spin
mesons (the data are, however, sparse). It would be certainly
interesting and important to observe systematically multiplets
of parity-chiral and parity-$U(1)_A$ groups  (or, sometimes, when
the chiral and $U(1)_A$ transformations connect {\it different}
hadrons \cite{G2}, the multiplets of the
$U(2)_L \times U(2)_R$ group). The high-lying hadron spectra
must be systematically explored.
This experimental task is just for existing facilities like
JLAB, BNL, SPRING 8, ELSA, as well as for the forthcoming Japanese hadron
facility and the proton-antiproton ring in Darmstadt.\\

The strength of the argument given above is that it is very
general. Its weakness is that we
cannot say anything concrete about microscopical mechanisms of
how all this happens. For that one needs a detailed microscopical
understanding of dynamics in QCD, which is both challenging and
very difficult task. But even though we do not know 
how microscopically all this happens, we can claim that in highly excited
hadrons we must observe symmetries of the classical QCD Lagrangian.
The only basis for this statement is that in such hadrons a semiclassical
description is correct. \\

As a consequence, in highly excited hadrons the valence quark
motion has to be described semiclassically and at the same time
their chirality (helicity) must be fixed. Also the gluonic
field should be described semiclassically. All this gives an
increasing support for a string picture of highly excited hadrons,
where hadrons are viewed as strings with massless quarks of
definite chirality at the end-points of the string \cite{G3}.\\

\bigskip

The work was supported by the FWF project P14806-TPH
of the Austrian Science Fund.


\bigskip



\begin{references}
\bibitem{ANOMALY} S. L. Adler, Phys. Rev. {\bf D}, 2426 (1969);
J. S. Bell and R. Jaiw, Nuovo Cim. {\bf 60A}, 47 (1969).
\bibitem{CG1} T. D. Cohen and L. Ya. Glozman, Phys. Rev. 
{\bf D65},016006 (2002); Int. J. Mod. Phys. {\bf A17},  1327 (2002).
\bibitem{W} S. Weinberg, Phys. Rev. Lett. {\bf 166}, 1568 (1968).
\bibitem{PDG} Particle Data Group, D. Groom et al, Eur. Phys. J.
{\bf C15}, 1 (2000).
\bibitem{G1} L. Ya. Glozman, Phys. Lett. {\bf B475}, 329 (2000).
\bibitem{ERICE} L. Ya. Glozman, hep-ph/0210216.
\bibitem{BUGG1} A. V. Anisovich et al, Phys. Lett. {\bf B491}, 47 (2000).
\bibitem{BUGG2} A. V. Anisovich et al, Phys. Lett. {\bf B517}, 261 (2001).
\bibitem{BUGG0} A. V. Anisovich et al, Phys. Lett. {\bf B449},154 (1999);
 BES Collaboration , Phys. Lett. {\bf B472}, 207 (2000).
\bibitem{G2}L. Ya. Glozman, Phys. Lett. {\bf B539}, 257 (2002).
\bibitem{G3}L. Ya. Glozman, Phys. Lett. {\bf B541}, 115 (2002).
\bibitem{SVZ} M. A. Shifman, A. I. Vainstein, V. I. Zakharov,
Nucl. Phys. {\bf B147}, 385 (1979).
\bibitem{SS} T. Sch\"afer and E. Shuryak, Rev.Mod. Phys.
{\bf 70}, 323 (1998); D. Diakonov, hep-ph/0212026.
\bibitem{RWA} C. D. Roberts and A. G. Williams, Progr. Part. Nucl. Phys.,
{\bf 33}, 477 (1994); R. Alkofer and L. von Smekal, Phys. Rept.
{bf 353}, 281 (2001).

\end{references}
\end{document}